\documentclass[aps,prl,reprint,amssymb,superscriptaddress]{revtex4-1}

\usepackage{amsmath,amssymb,color}
\usepackage{bm}
\usepackage{hyperref}
\usepackage{graphicx}

\begin{document}

\title{Bulk and edge spin transport in topological magnon insulators}

\author{Andreas R\"{u}ckriegel}
\affiliation{Institute for Theoretical Physics and Center for Extreme Matter and Emergent Phenomena,
Utrecht University, Leuvenlaan 4, 3584 CE Utrecht, The Netherlands}

\author{Arne Brataas}
\affiliation{Center for Quantum Spintronics, Department of Physics, Norwegian University of Science and Technology, 
NO-7491 Trondheim, Norway}

\author{Rembert A. Duine}
\affiliation{Institute for Theoretical Physics and Center for Extreme Matter and Emergent Phenomena,
Utrecht University, Leuvenlaan 4, 3584 CE Utrecht, The Netherlands}
\affiliation{Center for Quantum Spintronics, Department of Physics, Norwegian University of Science and Technology, 
NO-7491 Trondheim, Norway}
\affiliation{Department of Applied Physics, Eindhoven University of Technology,
P.O. Box 513, 5600 MB Eindhoven, The Netherlands}

\date{November 3, 2017}

\begin{abstract}
We investigate the spin transport properties of a topological magnon insulator,
a magnetic insulator
characterized by topologically nontrivial bulk magnon bands and protected magnon edge modes located in the bulk band gaps.
Employing the Landau-Lifshitz-Gilbert phenomenology,
we calculate the spin current driven through a 
normal metal$|$topological magnon insulator$|$normal metal heterostructure by a spin accumulation imbalance between the metals,
with and without random lattice defects.
We show that bulk and edge transport are characterized by different length scales. 
This results in a characteristic system size where the magnon transport crosses over from being bulk-dominated for small systems to edge-dominated for larger systems.
These findings are generic and relevant for topological transport in systems of nonconserved bosons.
%In clean systems, Gilbert damping affects the edge current more strongly than the bulk current.
%In contrast, in dirty samples,
%topologically protected edge magnons dominate long-range spin transport.
\end{abstract}

\maketitle

\textit{Introduction.$-$}
Topological insulators are electronic states of matter 
where the bulk insulates and the edge conducts.
The edge modes are protected against weak perturbations and backscattering by the topology of the bulk band structure \cite{Hasan10}.
They are also chiral and propagate unidirectionally along the sample boundaries.
 
The notion of topologically protected edge states is not unique to electronic systems.
A growing number of studies investigates the possibility of topological magnon insulators (TMIs) \cite{Zhang13,Shindou13a,Shindou13b,Mook14,Cao15,Chisnell15,Owerre16a,Owerre16b,Owerre16c,Kim16,Mook16,Xu16,Wang17,Laurell17,Owerre17a,Owerre17b,Owerre17c,Owerre17d,Owerre17e,Owerre17f,Nakata17a,Nakata17b}
as well as magnon Dirac \cite{Owerre17g,Okuma17,Li17,Yao17,Bao17} and Weyl semimetals \cite{Li16b,Mook16b,Mook17,Su17a,Su17b,Owerre17h}.
Magnons are collective excitations in magnetic systems in which the spins precess coherently around the direction of the local magnetic order.
As they allow one to transport spin with low dissipation over long distances \cite{Cornelissen15},  
magnon-based devices are promising for the growing field of spintronics \cite{Zutic04}.
In TMIs, the magnon band structure exhibits a topologically nontrivial energy gap 
that hosts protected edge states.
These edge modes open a new channel for spin transport.
Their robustness to perturbations as well as their chiral nature 
makes them appealing from the perspective of applications. 
There is an important difference between electronic and magnonic topological insulators
arising from their difference in quantum statistics:
While edge states in electronic systems are gapless low-energy excitations,
the edge modes in TMIs are high-energy states with a sizable gap.
In ferromagnets,
the edge modes therefore coexist with low-energy bulk modes, 
and both will contribute to the spin transport \cite{Note1}.

While much effort has been devoted to identifying possible candidates for TMIs \cite{Zhang13,Shindou13a,Shindou13b,Cao15,Chisnell15,Owerre16a,Owerre16c,Kim16,Owerre17a,Owerre17b,Owerre17c,Owerre17d,Owerre17e,Nakata17a,Nakata17b},
the transport properties of TMIs have received considerably less attention. 
Most studies so far focus on the thermal Hall response of ideal TMIs,
disregarding the ubiquitous sources of dissipation present in materials.
In particular, magnetic insulators suffer from Gilbert damping,
which needs to be included for a realistic description of magnon spin transport.
Recent advances in electrical spin injection in normal metal$|$ferromagnet heterostructures and long-range spin transport \cite{Cornelissen15,Goennenwein15,Wu16,Li16,Cornelissen16} 
motivate a study of
the relative contribution of the edge states to the longitudinal spin transport,
as well as of the response to electrical rather than thermal driving forces.

\begin{figure}
\includegraphics[width=.45\textwidth]{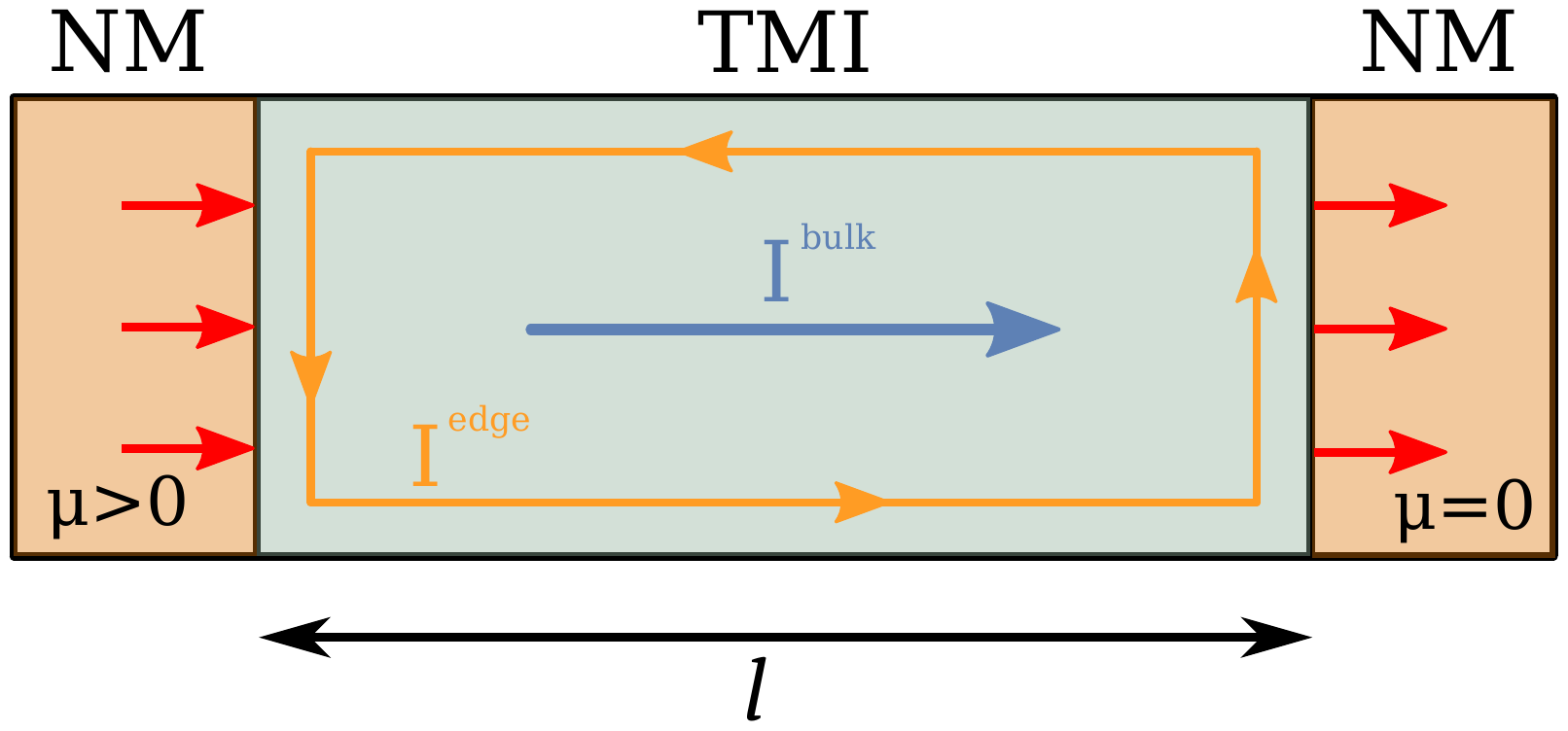}
\caption{\label{fig1}Spin transport in a topological magnon insulator (TMI) of length $l$,
driven by the difference in spin accumulation $\bm{\mu}$ in the attached normal metal (NM) leads.
Apart from the spin current $I^{\textrm{bulk}}$ carried by bulk magnons,
in the TMI there is also a current $I^{\textrm{edge}}$ 
carried by topologically protected edge excitations.
This edge current circles the system boundaries in a chiral fashion 
and is robust to perturbations \cite{Hasan10}.} 
\end{figure}
In this Rapid Communication,
we use the stochastic Landau-Lifshitz-Gilbert formalism \cite{Hoffmann13,Brataas15,Bender17,Zheng17}
to address these open questions.
We consider diffusive spin transport through a 2-dimensional normal metal$|$TMI$|$normal metal heterostructure,
driven by a difference in spin accumulation between the metals, see Fig.~\ref{fig1}.
Within linear spin wave theory,
we numerically calculate the spin current carried by both bulk and edge magnon modes,
and find that they are described by two different spin diffusion lengths.
We furthermore investigate
the influence of Gilbert damping, temperature and random lattice defects on the edge and bulk spin currents.
Our findings are that in a clean system the edge current is strongly suppressed by Gilbert damping 
and the large excitation gap.
On the other hand, 
adding random defects strongly diminishes the bulk current while having only a weak effect on the edge currents,
reflecting the topological protection of the edge states.    
For disordered systems we thus find that the edge states dominate the spin transport for sufficiently large systems.
While focusing on magnons, 
we believe our results to be generic and to apply to physical situations where topological transport is carried by nonconserved bosons, such as e.g. certain photonic crystals \cite{Raghu08,Wang09,Lu14}.

\begin{figure} 
\includegraphics[width=.48\textwidth]{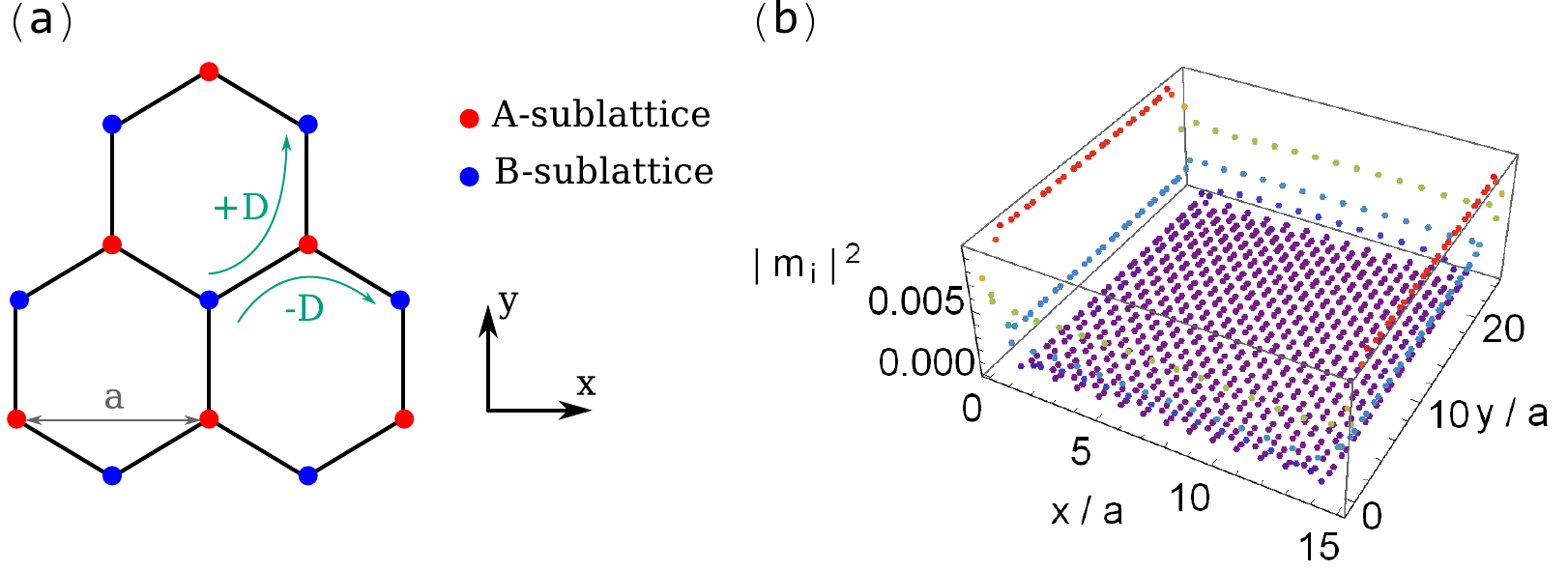}
\caption{\label{fig2}(a) The honeycomb lattice structure, 
consisting of two interpenetrating triangular sublattices with lattice constant $a$.
The DMI changes sign depending on whether there is a right or left turn 
between next-nearest neighbors.
(b) A typical magnon edge state with frequency $\omega=2.73\,J$ in the bulk band gap, obtained by diagonalizing the inverse magnon propagator (\ref{eq:G_inverse}) for vanishing damping and spin pumping ($\alpha_i=0=\mu_i$) on a finite lattice with $930$ lattice sites.
Spin, magnetic field and DMI are fixed to $S=1$, $H=0.1\,J$ and $D=0.2\,J$ in the plot.}
\end{figure}

\textit{Model and formalism.$-$}
We consider a TMI described by the Hamiltonian
\begin{equation} \label{eq:H}
\mathcal{H} = 
\frac{1}{2} \sum_{ij} \Bigl[
-J_{ij} \bm{S}_i \cdot \bm{S}_j
+D_{ij} \bm{\hat{z}} \cdot \left( \bm{S}_i \times \bm{S}_j \right)
\Bigr]
-H \sum_i S_i^z ,
\end{equation}
where the $\bm{S}_i$ are spins of magnitude $S$ localized at the sites $\bm{R}_i$ of a 2-dimensional honeycomb lattice in the $x$-$y$ plane,
see Fig.~\ref{fig2} (a). 
The exchange coupling is $J_{ij}=J>0$ for nearest neighbors, 
and $D_{ij}=-D_{ji}=D>0$ is the Dzyaloshinskii-Moriya interaction (DMI) between next-nearest neighbors.
Lastly, $H$ is the Zeeman energy associated with an external magnetic field applied in the $z$ direction that stabilizes the ferromagnetic ground state.
In the ordered phase, 
the Hamiltonian (\ref{eq:H}) is known to be the bosonic analog of the Haldane model \cite{Kim16,Haldane88}.
When the DMI is finite, the system exhibits topologically nontrivial magnon bands 
and corresponding protected edge states.

Spin dynamics are governed by the stochastic Landau-Lifshitz-Gilbert equation (LLG):
\begin{equation} \label{eq:LLG}
\partial_t \bm{S}_i 
= \bm{S}_i \times \left[   
-\frac{ \partial \mathcal{H} }{ \partial \bm{S}_i } + \bm{h}_i(t)
- \frac{\alpha_i}{S} \partial_t \bm{S}_i 
+ \frac{\alpha_i^{\textrm{sp}}}{S} \bm{S}_i \times \bm{\mu}_i
\right]  
\end{equation}
where the Gilbert damping $\alpha_i = \alpha + \alpha_i^{\textrm{sp}}$
incorporates the bulk Gilbert damping $\alpha$ 
as well as the interface Gilbert damping enhancement $\alpha_i^{\textrm{sp}}$ \cite{Tserkovnyak02}, 
and $\bm{\mu}_i=\mu_i\bm{\hat{z}}$ is the spin accumulation in the left lead 
established, e.g., by the spin Hall effect.
Note that since both Gilbert damping enhancement $\alpha_i^{\textrm{sp}}$ and spin accumulation $\mu_i$
are only finite on lattice sites connected to the leads,
the corresponding spin-transfer torque terms in the LLG (\ref{eq:LLG})
are boundary conditions.
The stochastic magnetic field $\bm{h}_i(t)$ models thermal fluctuations in the TMI
and will be fixed by the fluctuation-dissipation theorem (FDT) in Eq.~(\ref{eq:FDT}) below.

When $H\ge 0$ and $D<J/\sqrt{3}$, 
the bulk ground state of the Hamiltonian (\ref{eq:H}) is the uniform state $\bm{S}_i = S\bm{\hat{z}}$.
Focusing on this case, 
we linearize in the deviations $m_i=( S_i^x + i S_i^y )/S$ from this uniform state 
to obtain the magnon equation of motion, 
which becomes in frequency space
\begin{equation} \label{eq:LLG_linear}
\sum_j \mathbb{G}^{-1}_{ij}(\omega) m_j(\omega) = h_i(\omega) .
\end{equation}
Here, $h_i(\omega)$ is the Fourier transform of the circular component $h_i=h_i^x + i h_i^y$ of the stochastic field,
and the matrix elements of the inverse magnon propagator are given by  
\begin{equation} \label{eq:G_inverse}
\mathbb{G}^{-1}_{ij}(\omega) = 
\delta_{ij} \Bigl[ 
-\left( 1 + i \alpha_i \right) \omega 
+ H + S \sum_n J_{in} + i \alpha_i^{\textrm{sp}} \mu_i
\Bigr] 
+ \mathbb{T}_{ij} ,
\end{equation}
with the complex hopping matrix elements $\mathbb{T}_{ij} = -S ( J_{ij}+iD_{ij} )$.
The magnon spectrum and eigenstates can be obtained by 
diagonalizing the inverse propagator (\ref{eq:G_inverse}) in the absence of damping and spin pumping.
A typical magnon edge state with frequency in the bulk band gap is displayed in Fig.~\ref{fig2} (b). 
Explicit calculations of the bulk band structure, Berry curvature, and associated Chern numbers are reviewed in the Supplemental Material \cite{Supplement}.

At finite temperatures the magnon spectrum is populated by thermal fluctuations 
encoded in the stochastic magnetic field $h_i(\omega)$.
To ensure agreement with the equilibrium predicted by the quantum-mechanical linear spin-wave theory for magnons, 
we set $\langle h_i(\omega) \rangle = 0$,    
and $\langle h_i(\omega) h_j^*(\omega') \rangle =
2\pi \delta(\omega-\omega') \mathbb{R}_{ij}(\omega)$,
with a covariance matrix determined by the quantum-mechanical FDT \cite{Brataas15,Bender17,Zheng17,Note2}: 
\begin{equation} \label{eq:FDT}
\mathbb{R}_{ij}(\omega) =\delta_{ij}
\frac{4\alpha_i(\omega-\mu_i)/S}{ e^{(\omega-\mu_i)/k_B T} - 1 } . 
\end{equation}

Finally,
the total spin current $I$ ejected into the right lead in a stationary state can be obtained from 
$\partial_t \langle S_i^z \rangle = 0$ (see Supplemental Material \cite{Supplement} for details)
and can be written as 
$I = \int \frac{\textrm{d}\omega}{2\pi} I(\omega)$,
with
\begin{equation} \label{eq:I_omega}
I(\omega) = \sum_{i\in\textrm{interface}} \textrm{Im}
\Bigl[ S \mathbb{T} \mathbb{G}(\omega) \mathbb{R}(\omega) \mathbb{G}^\dagger(\omega)  \Bigr]_{ii} ,
\end{equation}
where the sum is over all lattice sites in contact with the right lead.
By splitting the frequency integration into integrals over the bulk bands and the bulk gap, 
$\int \frac{\textrm{d}\omega}{2\pi}=
\int_{\textrm{bulk}} \frac{\textrm{d}\omega}{2\pi} + 
\int_{\textrm{gap}} \frac{\textrm{d}\omega}{2\pi}$,
we can separate the contributions from the bulk bands and the edge states to the spin current,
$I = I^{\textrm{bulk}} + I^{\textrm{edge}}$,
and analyze them separately \cite{NoteSeparation}.

\textit{Numerical results and discussion.$-$}
In the following, we will present numerical results for the spin current $I$ ejected into the right lead.
We consider a finite 2 dimensional honeycomb lattice with zigzag termination on the sides attached to the leads
and armchair termination on the other two boundaries.
The system has a variable length $l$ and a fixed width of $11$ lattice sites
that is larger than the penetration width of the edge states.
Only the outermost lattice sites of the zigzag edges are connected to the leads.
The spin, applied field and DMI will be fixed 
to $S=1$, $H=0.1\,J$, and $D=0.2\,J$ from here on.
For this set of parameters, 
the edge state gap predicted by the bulk band structure is $\Delta = 2.1\,J$.
Gilbert damping enhancement and spin accumulation are set to 
$\alpha_i^{\textrm{sp}}=1$ and $\mu_i=0.01\,J$ respectively.

For the numerical solution,
we have obtained the magnon propagator as a function of frequency 
by direct inversion of Eq.~(\ref{eq:G_inverse}).
The bulk and edge spin currents,
$I^\textrm{bulk}$ and $I^\textrm{edge}$, 
are subsequently calculated 
by integrating Eq.~(\ref{eq:I_omega}) over the respective frequency ranges. 
To study the effect of lattice defects on the spin currents,
a large on-site field is added to randomly chosen lattice sites,
effectively making them inaccessible to magnons.
The lattice sites are chosen with a probability $w$
and the resultant spin currents are averaged over many realizations of defect distributions,
so that the average defect concentration is $w$. 
In practice, averaging over $25$ defect realizations is sufficient for convergence.

\begin{figure}
\includegraphics[width=.48\textwidth]{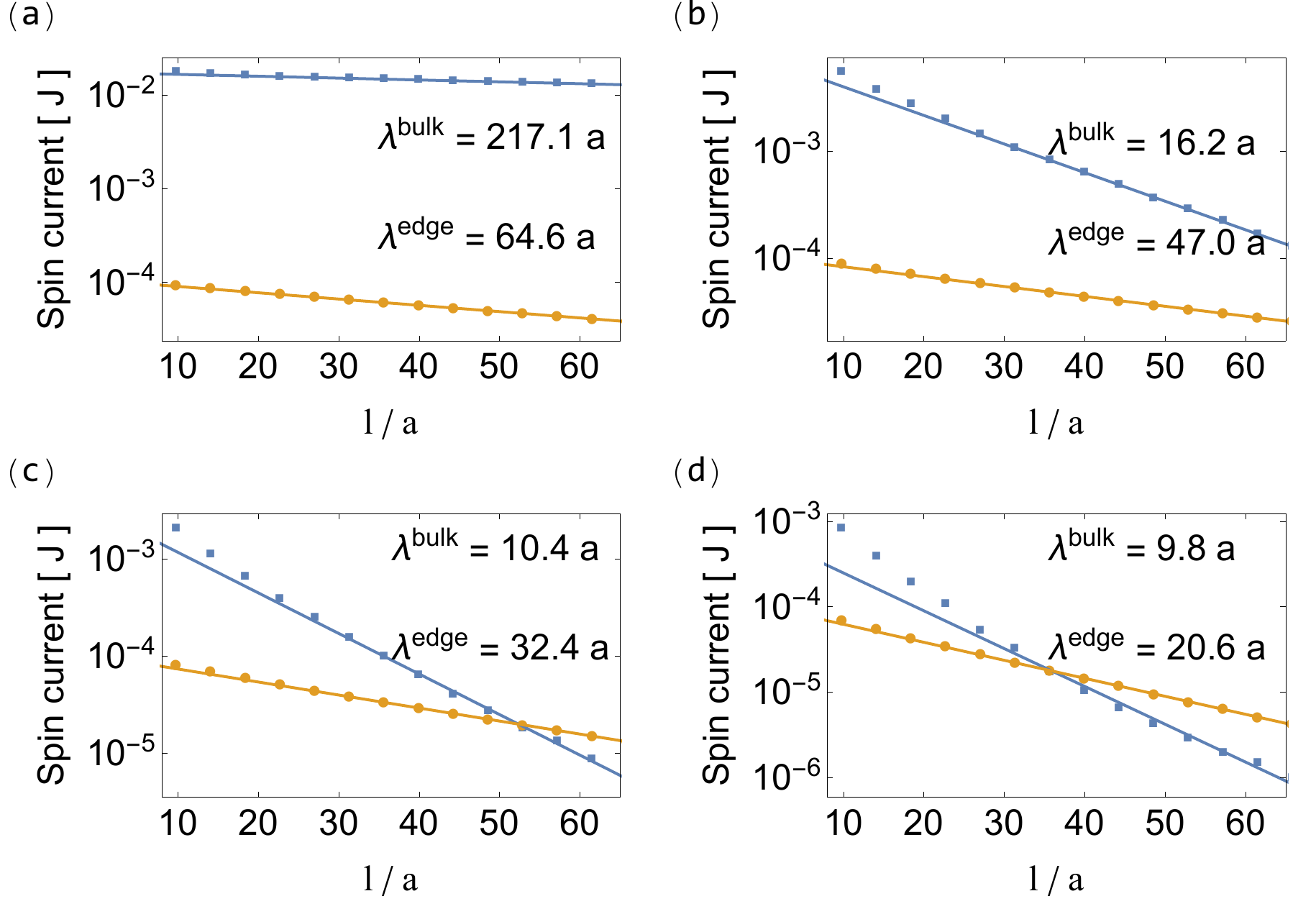}
\caption{\label{fig3}Spin currents ejected at the right lead as a function of the system length $l$,
for temperature $T=0.8\,J$, bulk Gilbert damping $\alpha=2.5\times 10^{-3}$, and different average defect concentrations $w$.
Bulk and edge contributions are depicted as blue squares and yellow circles respectively.
The corresponding straight lines are exponential fits according to Eq. (\ref{eq:I_exp}),
with the spin diffusion lengths denoted in the plots. 
The defect concentrations are
(a) $w=0$ (clean system),
(b) $w=0.05$,
(c) $w=0.1$,
(d) $w=0.15$.
Deviations from the exponential fit for $l\alt 25\,a$ signal the crossover to the thin-film regime
in which the spin currents decay algebraically \cite{Cornelissen15,Cornelissen16}.}
\end{figure}
Figure \ref{fig3} shows the dependence of the ejected spin current on the length $l$ of the  sample
for different average defect concentrations $w$.
For each $w$,
the spin currents decays exponentially with increasing length $l$ \cite{Note3},   
\begin{equation} \label{eq:I_exp}
I^X(l) \propto \exp\left( -l/\lambda^X \right),
\,\,\,
X = \textrm{bulk}, \textrm{edge} ,
\end{equation}
with respective spin diffusion lengths $\lambda^\textrm{bulk}$ and $\lambda^\textrm{edge}$.
In the clean limit, $w=0$,
the bulk contribution to the spin currents decays far slower than the edge contribution,
see Fig.~\ref{fig3} (a).
This is expected because the only relaxation mechanism in this case is the Gilbert damping,
which is proportional to the frequency of the magnon.
As the edge states are high-frequency states in the bulk band gap, 
they are affected far stronger by Gilbert damping than the low-frequency
bulk magnons supporting the bulk spin current.
At the same time, 
the total spin current carried by the edge magnons is 2 to 3 orders of magnitude smaller than the bulk current, 
due to the exponential suppression of the high-frequency edge states by the thermal Bose distribution in the FDT (\ref{eq:FDT}).
However, as shown in Figs.~\ref{fig3} (b)-(d),
adding defects has a dramatic effect on this:
while both bulk and edge diffusion lengths decrease,
the effect on the bulk is far stronger,
so that eventually the bulk contribution to the spin current decays,
leaving only the edge current. 
A discussion of the dependence of the spin diffusion lengths on the bulk Gilbert damping is relegated to the Supplemental Material \cite{Supplement}.

\begin{figure}
\includegraphics[width=.45\textwidth]{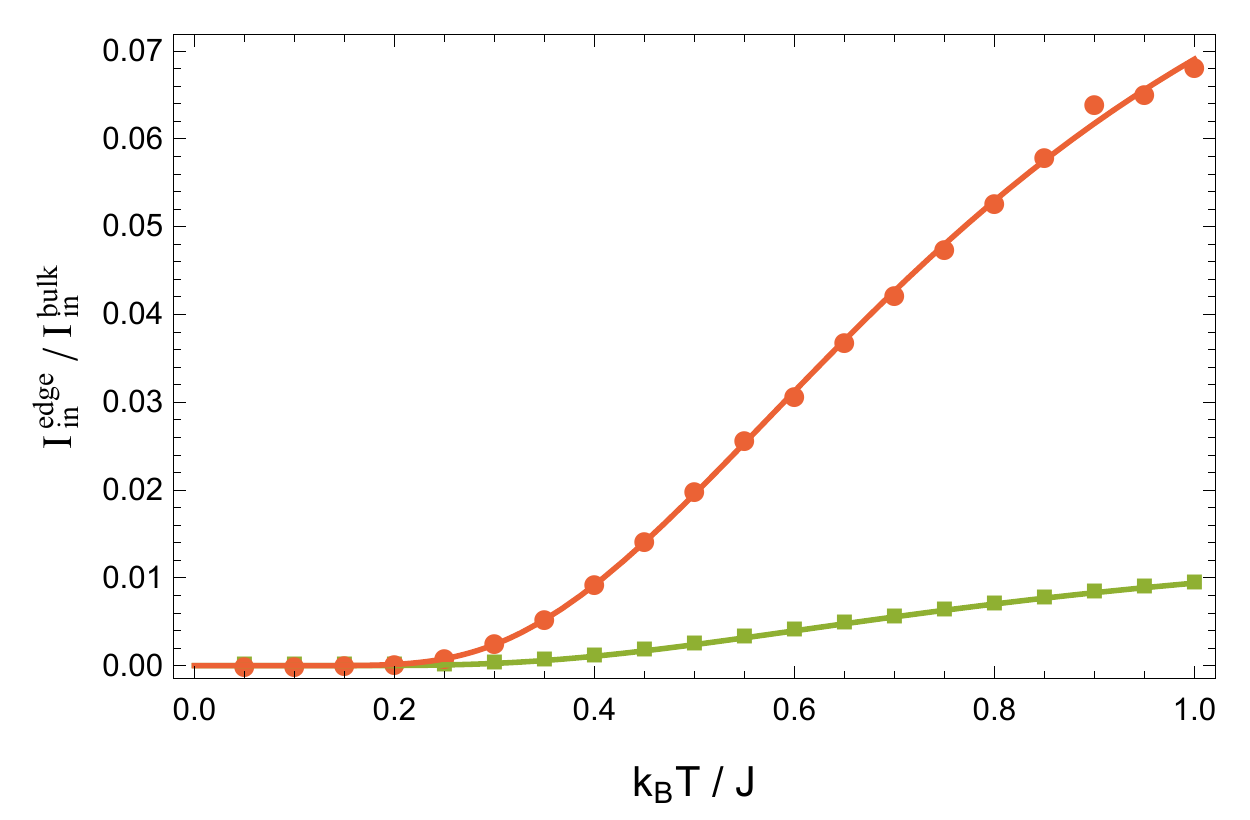}
\caption{\label{fig4}Ratio of edge to bulk current injected at the left lead as a function of temperature $T$,
for bulk Gilbert damping $\alpha=5\times 10^{-3}$ and average defect concentrations $w=0$ (green squares) and $w=0.1$ (red circles).
The corresponding lines are fits with Eq.~(\ref{eq:Ratio}).} 
\end{figure}

In Fig.~\ref{fig4} we plot the ratio of the edge to the bulk spin current injected into the TMI at the left lead as a function of temperature. 
As anticipated, this ratio can be fitted to an exponential \cite{Note4}:  
\begin{equation} \label{eq:Ratio}
\frac{I^{\textrm{edge}}_{\textrm{in}}(T)}{I^{\textrm{bulk}}_{\textrm{in}}(T)} \propto 
\frac{J}{k_B T} \exp\left( -\nu \frac{J}{k_B T} \right) ,
\end{equation}
with a constant $\nu$ that is found to be close to $2$ both with and without disorder.  
This reflects the exponential suppression of the edge states by 
their excitation gap $\Delta=2.1\,J$.
The additional prefactor of $1/T$ stems from the bulk spin current 
which is dominated by the thermally populated low-frequency magnons.
The ratio of the injected currents significantly increases by adding defects.
This is caused by the weak localization of the bulk magnon states induced by the disorder,
which decreases the conductivity of the bulk and is a precursor to Anderson localization \cite{Evers15,Evers17}.
On the other hand, the edge states are protected by the topology of the system and are only weakly affected by the random defects, 
compare also Fig.~\ref{fig3}.   

\begin{figure}
\includegraphics[width=.45\textwidth]{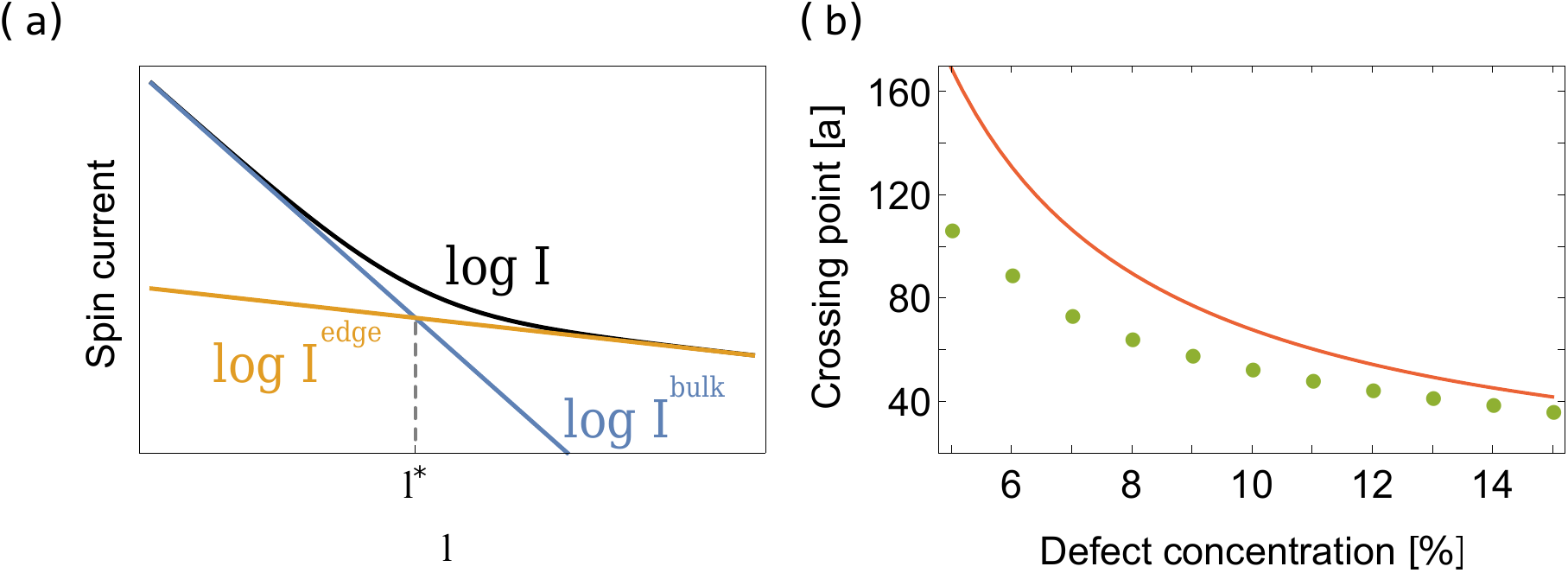}
\caption{\label{fig5}(a) Schematic depiction of the contribution of edge and bulk magnons to the total spin current, 
with the crossover length scale $l^*$.
Note that $l^*\to\infty$ for clean systems, see Fig.~\ref{fig3} (a).
(b) Numerically calculated crossing points $l^*$ (points) and the order of magnitude estimate presented in the text (solid line).
Parameters are the same as in Fig.~\ref{fig3}.} 
\end{figure}

Lastly, we give a simple order of magnitude estimate of the crossing length scale $l^*$ at which the edge current overtakes the bulk current,
depicted schematically in Fig.~\ref{fig5} (a).
Assuming an exponential decay as in Eq.~(\ref{eq:I_exp}) throughout,
we have 
$l^* = \ln( I^{\textrm{bulk}}_{\textrm{in}} / I^{\textrm{edge}}_{\textrm{in}} ) / 
( 1/\lambda^{\textrm{bulk}} - 1/\lambda^{\textrm{edge}} )$.
The injected currents may be estimated as $I^{\textrm{bulk}}_{\textrm{in}} \sim k_B T /H $
and $I^{\textrm{edge}}_{\textrm{in}} \sim \exp(-\Delta /k_B T)$,
where $H$ is the bottom of the bulk magnon spectrum; 
see Eq.~(\ref{eq:Ratio}) and the subsequent discussion.
The bulk spin diffusion length is dominated by the disorder, see Fig.~\ref{fig3},
therefore we assume 
$\lambda^{\textrm{bulk}}\sim l_{\textrm{imp}}\sim a/w$,
where $l_{\textrm{imp}}$ is the mean free path due to defect scattering.
The spin diffusion length of the edge magnons is estimated as 
$\lambda^{\textrm{edge}} = v\sqrt{\frac{2}{3} \tau_{\textrm{r}} \tau}$ \cite{Cornelissen16,Supplement}.
Here, $v=2\sqrt{J k_B T}a$ is the average magnon velocity,
$\tau_{\textrm{r}}^{-1}=1/(2\alpha \Delta )$
is the magnon relaxation rate due to inelastic scattering,
and the total magnon relaxation rate is
$\tau^{-1} = \tau_{\textrm{r}}^{-1} + \tau_{\textrm{el}}^{-1}$
and also includes elastic scattering with defects and the sample boundaries
taken into account by $\tau_{\textrm{el}}^{-1}=l_{\textrm{imp}}/v$. 
%with average magnon velocity $v=2\sqrt{J k_B T}a$,
%and relaxation times $\tau_{\textrm{r}}=1/(2\alpha \Delta )$
%and $\tau_{\textrm{el}}=l_{\textrm{imp}}/v$.
Although we ignore both the thin-film regime and the effect of Gilbert damping on the bulk spin current,
Fig.~\ref{fig5} (b) shows a reasonable agreement of our estimate with the actual numerical results,
especially for higher defect concentrations where the aforementioned effects are less important.

\textit{Conclusions.$-$}
Edge states protected by topology are a promising new tool for spintronics applications. 
Although the edge spin current is strongly suppressed by the gap of the edge magnons,
we have shown that in a disordered system the protected edge states dominate the long-distance spin transport. 
While we have studied a particular model system, the honeycomb ferromagnet,
we believe that our results are generic and pertain to all ferromagnetic topological magnon insulators
and, more generally, to topological boson systems.
In particular, our theory should apply to the recently discovered topological magnon insulator on a kagome lattice \cite{Chisnell15},
as well as to the proposed magnonic crystals with topologically nontrivial magnon bands \cite{Shindou13a}.
The honeycomb ferromagnet with Dzyaloshinskii-Moriya interaction that we investigated may also be realized experimentally
by depositing magnetic impurities on a metal with strong spin-orbit coupling \cite{Kim16,Fransson16}.
The latter two proposals have the additional advantage that the amount of disorder can be experimentally controlled.

To directly connect to possible experiments, let us estimate the crossover length scale $l^*$ for two representative examples and moderate disorder concentrations $w\sim 0.1$.
For the kagome system Cu(1,3-bdc) investigated in \cite{Chisnell15}, we find $l^* \sim 0.1\,\mu \textrm{m}$ for $T\sim 1\,\textrm{K}$, $H\sim 1\,\textrm{T}$, and assuming $\alpha\sim 10^{-2}$. 
This is roughly an order of magnitude larger than the estimates we obtain for the spin diffusion lengths.
We thus expect this system to be in the regime where bulk magnons dominate the spin transport.
On the other hand, for the YIG$|$Fe magnonic crystal envisioned in \cite{Shindou13a}, we estimate a crossover length scale $l^* \sim 0.1\,\mu\textrm{m}$ at room temperature, 
while the bulk spin diffusion length is $\sim 1\,\mu\textrm{m}$. Therefore spin currents carried mainly by edge magnons should be readily observable in this system.
 
Future work should be focused on a more microscopic modeling of the damping of the edge modes beyond the Landau-Lifshitz-Gilbert paradigm. 
Especially at elevated temperatures when the density of bulk magnons is large,
spin-wave interactions between bulk and edge magnons
may also change the transport properties of both  drastically.

\begin{acknowledgments}{\textit{Acknowledgments.$-$}}
RD acknowledges support as a member of the D-ITP consortium, a program of the Netherlands Organisation for Scientific Research (NWO) that is funded by the Dutch Ministry of Education, Culture and Science (OCW). This work is in part funded by the Stichting voor Fundamenteel Onderzoek der Materie (FOM). 
This project has received funding from the European Research Council (ERC) under the European Union's 
Horizon 2020 research and innovation programme (grant agreement 725509 - SPINBEYOND),
and was partially supported by the Research Council of Norway through its Centres of Excellence funding scheme, project number $262633$, ``QuSpin".
\end{acknowledgments}

\pagebreak

%%%%%%%%%% Prefix a "S" to all equations, figures, tables and reset the counter %%%%%%%%%%
\setcounter{equation}{0}
\setcounter{figure}{0}
\setcounter{table}{0}
\setcounter{page}{1}
\makeatletter
\renewcommand{\theequation}{S\arabic{equation}}
\renewcommand{\thefigure}{S\arabic{figure}}
\renewcommand{\theHfigure}{S.\thefigure}
\renewcommand{\bibnumfmt}[1]{[S#1]}
\renewcommand{\citenumfont}[1]{S#1}
%%%%%%%%%% Prefix a "S" to all equations, figures, tables and reset the counter %%%%%%%%%%

\begin{widetext}

\section{Supplemental Material}

\appendix

\section{Bulk bands, Berry curvature, and Chern numbers}

In order to obtain the magnon band structure in a bulk system,
we may neglect the effects of Gilbert damping as well as spin pumping,
and impose periodic boundary conditions,
so that the magnon equation of motion (\ref{eq:LLG_linear}) becomes
\begin{equation} \label{eq:bulk}
i\partial_t m_i = \left( H + 3SJ \right) m_i -S \sum_j \left( J_{ij} + i D_{ij} \right) m_j .
\end{equation}
Due to the periodic boundary conditions in the bulk, 
it is advantageous to Fourier transform to momentum space via
\begin{equation}
m_i = \sqrt{\frac{2}{N}} \sum_{\bm{k}} e^{i\bm{k}\cdot\bm{R}_i} \times
\begin{cases}
a_{\bm{k}} & \bm{R}_i \in \mathcal{A} \\
b_{\bm{k}} & \bm{R}_i \in \mathcal{B}
\end{cases} ,
\end{equation}
where $N$ is the total number of lattice sites,
and $\mathcal{A}$ and $\mathcal{B}$ denote the two sublattices of the honeycomb lattice, 
see Fig.~\ref{fig2} (a).
This turns the above Eq.~(\ref{eq:bulk}) into 
\begin{equation} \label{eq:bulk_k}
i\partial_t \left( \begin{matrix} a_{\bm{k}} \\ b_{\bm{k}} \end{matrix} \right) =
\left[ \left( H+3SJ \right) I_2 + \bm{h}_{\bm{k}}\cdot\bm{\sigma} \right]
%\left( \begin{matrix}
%H + 3SJ + SD_{\bm{k}} & -SJ_{\bm{k}}^* \\
%-SJ_{\bm{k}} & H + 3SJ - SD_{\bm{k}}
%\end{matrix} \right)
\left( \begin{matrix} a_{\bm{k}} \\ b_{\bm{k}} \end{matrix} \right) ,
\end{equation}
where $I_2$ is the 2 dimensional identity matrix,
$\bm{\sigma}$ is the vector of Pauli matrices,
and
\begin{equation}
\bm{h}_{\bm{k}} = S \sum_{i=1}^3
\left( \begin{matrix}
-J \cos\left( \bm{k}\cdot\bm{\delta}_i \right) \\
J \sin\left( \bm{k}\cdot\bm{\delta}_i \right) \\
2D \sin\left( \bm{k}\cdot\bm{\rho}_i \right) \\
\end{matrix} \right) .
\end{equation}
Here the nearest neighbor vectors are defined as 
$\bm{\delta}_1=(0,-a/\sqrt{3})$,
$\bm{\delta}_{2,3}=(\pm a/2,a/\sqrt{3})$,
whereas the next-nearest neighbor vectors are
$\bm{\rho}_1=(a,0)$,
$\bm{\rho}_{2,3}=(-a/2,\pm\sqrt{3}a/2)$;
compare Fig.~\ref{fig2} (a).
From the equation of motion (\ref{eq:bulk_k}) we immediately obtain 
the dispersions of the two magnon branches,
\begin{equation} \label{eq:bulk_bands}
\omega_{\bm{k},\pm} = H + 3SJ \pm | \bm{h}_{\bm{k}} | .
\end{equation}
In the absence of DMI,
the dispersions of the lower and upper magnon branches touch at the corners of the Brillouin zone,
at the two Dirac points $\bm{K}\equiv(4\pi/3a,0)$ and $\bm{K}'\equiv(2\pi/3a,2\pi/\sqrt{3}a)$.
In the vicinity of these points, the magnon dispersions become linear \cite{Owerre16S}.
\begin{figure}
\includegraphics[width=.85\textwidth]{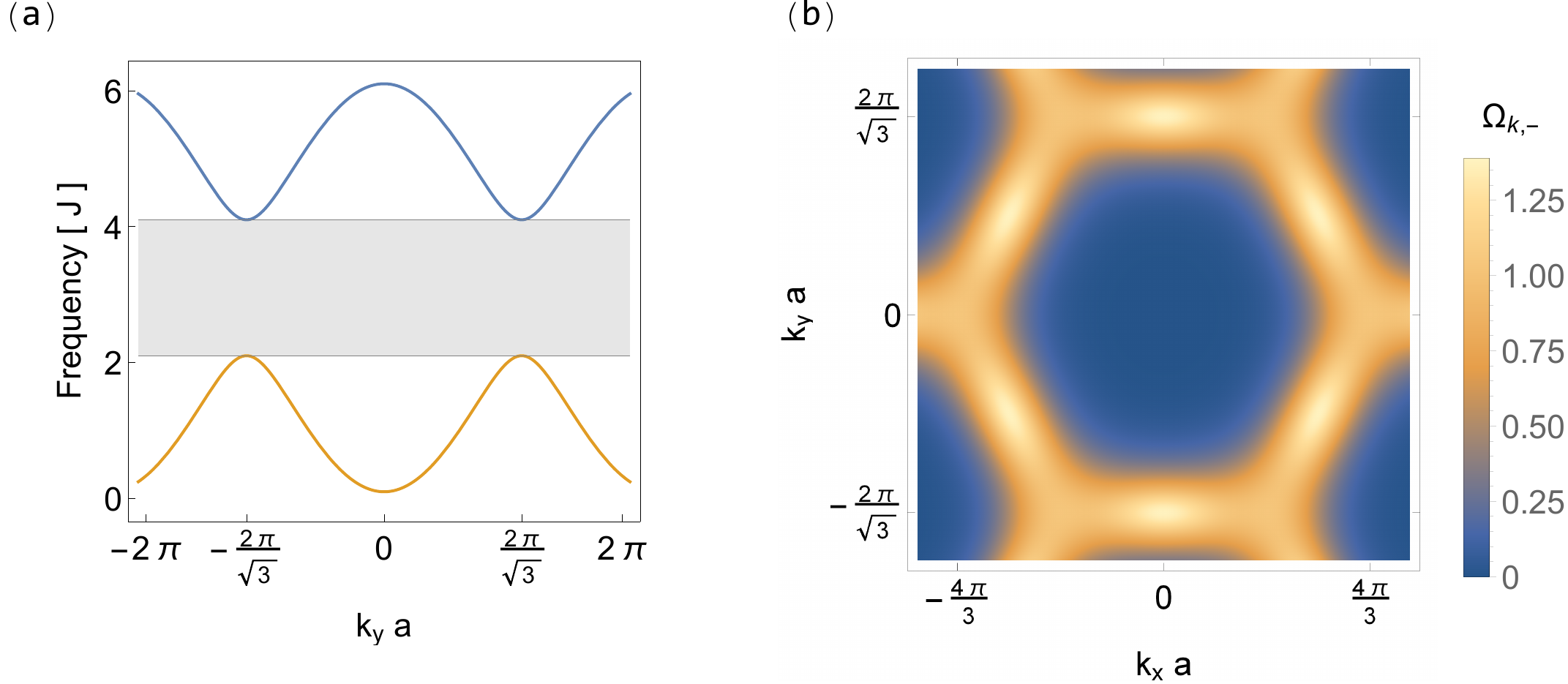}
\caption{\label{figS1}(a) Bulk magnon bands (\ref{eq:bulk_bands})
for momenta in $y$ direction. 
The shaded area denotes the bulk band gap.
(b) Intensity plot of the Berry curvature (\ref{eq:Berry}) of the lower magnon band.
Note that the Berry curvature is largest at the edges of the hexagonal Brillouin zone.
The Berry curvature of the upper magnon band is identical apart from a change of sign.
Both plots are for $S=1$, $H=0.1\,J$, and $D=0.2\,J$ as in the main text.} 
\end{figure}
Finite DMI however lifts this degeneracy and opens a gap,
thereby endowing the magnon bands with a nontrivial topology \cite{Owerre16S,Kim16S}.
A plot of the two magnon bands with finite DMI is displayed in Fig.~\ref{figS1} (a). 
The Berry curvatures of the magnon bands can be calculated from \cite{Kim16S}
\begin{equation} \label{eq:Berry}
\Omega_{\bm{k},\pm} = \mp \frac{\bm{\hat{h}}_{\bm{k}}}{2} \cdot 
\left( 
\frac{\partial \bm{\hat{h}}_{\bm{k}} }{ \partial k_x } \times
\frac{\partial \bm{\hat{h}}_{\bm{k}} }{ \partial k_y }
\right) ,
\end{equation}
where $\bm{\hat{h}}_{\bm{k}}= \bm{h}_{\bm{k}} / | \bm{h}_{\bm{k}} |$.
An intensity plot of the Berry curvature of the lower band is shown in Fig.~\ref{figS1} (b). 
The Chern numbers associated with the nontrivial topology of the two bands are
\begin{equation}
C_\pm = \frac{1}{2\pi} \int_{\textrm{BZ}} \textrm{d}^2k\, \Omega_{\bm{k},\pm} = \mp 1,
\end{equation}
where the integral is over the first Brillouin zone.

\section{Derivation of the spin current}

The spin current can in general be obtained from the equation of motion 
of the expectation value of the magnitude of the local magnetic moment \cite{Rueck17S}.
Since we are considering a ferromagnet that is fully polarized in $z$ direction,  
we therefore need the $z$ projection of the LLG (\ref{eq:LLG}),
given by
\begin{equation} \label{eq:continuity}
\partial_t S_i^z + \sum_j I_{i\to j} =
I_i^\alpha + I_i^h + I_i^\mu .
\end{equation}
The left-hand side of above Eq.~(\ref{eq:continuity}) 
has the form of a (lattice) continuity equation,
with hopping current
\begin{equation}
I_{i\to j} = S\, \textrm{Im} \left[ \mathbb{T}_{ij} m_i^* m_j \right] ,
\end{equation}
whereas the right-hand side consists of source terms
that describe the loss of spin to the lattice due to Gilbert damping ($\alpha$)
as well as the gain due to thermal fluctuations ($h$) and spin pumping ($\mu$),
\begin{subequations}
\begin{align}
I_i^\alpha =& -S \alpha_i\, \textrm{Im} \left[ m_i^* \partial_t m_i \right] , \\
I_i^h =& -S\, \textrm{Im} \left[ h_i^* m_i \right] , \\
I_i^\mu =& -S \alpha_i^{\textrm{sp}} \mu_i |m_i|^2 .
\end{align}
\end{subequations}
In a steady state,
we have $\partial_t \langle S_i^z \rangle = 0$,
so that the average spin loss (or gain) at each lattice site is compensated by the hopping current:  
\begin{equation}
I_i^{\textrm{loss}} \equiv \langle I_i^h + I_i^\alpha + I_i^\mu \rangle 
=\sum_j \langle I_{i\to j} \rangle .
\end{equation}
With the formal solution of the linearized LLG ($3$),
$m_i(\omega) = \sum_j \mathbb{G}_{ij}(\omega) h_j(\omega)$,
we thus find two equivalent expressions for the total spin loss at each lattice site:
\begin{align}
I_i^{\textrm{loss}}  
&= \int \frac{\textrm{d}\omega}{2\pi}\, S \left\{
\left( \alpha_i\omega - \alpha_i^{\textrm{sp}} \mu_i \right) 
\mathbb{G}(\omega) \mathbb{R}(\omega) \mathbb{G}^\dagger(\omega)
-\textrm{Im}\left[ \mathbb{G}(\omega) \right] \mathbb{R}(\omega)
\right\}_{ii} 
\\
&= \int \frac{\textrm{d}\omega}{2\pi}\, \textrm{Im}
\Bigl[ S \mathbb{T} \mathbb{G}(\omega) \mathbb{R}(\omega) \mathbb{G}^\dagger(\omega)  \Bigr]_{ii} .
\label{eq:loss_hopping}
\end{align}
The net spin current ejected or injected into the magnet at a given interface may
now be obtained as  
$I=\sum_{i\in \textrm{interface}} I_i^{\textrm{loss}} $.
Note that in the main text the second, more compact formulation for the loss current, Eq.~(\ref{eq:loss_hopping}), is used for convenience, compare Eq.~(\ref{eq:I_omega}).

\section{Damping dependence of the spin diffusion lengths}

\begin{figure}
\includegraphics[width=.85\textwidth]{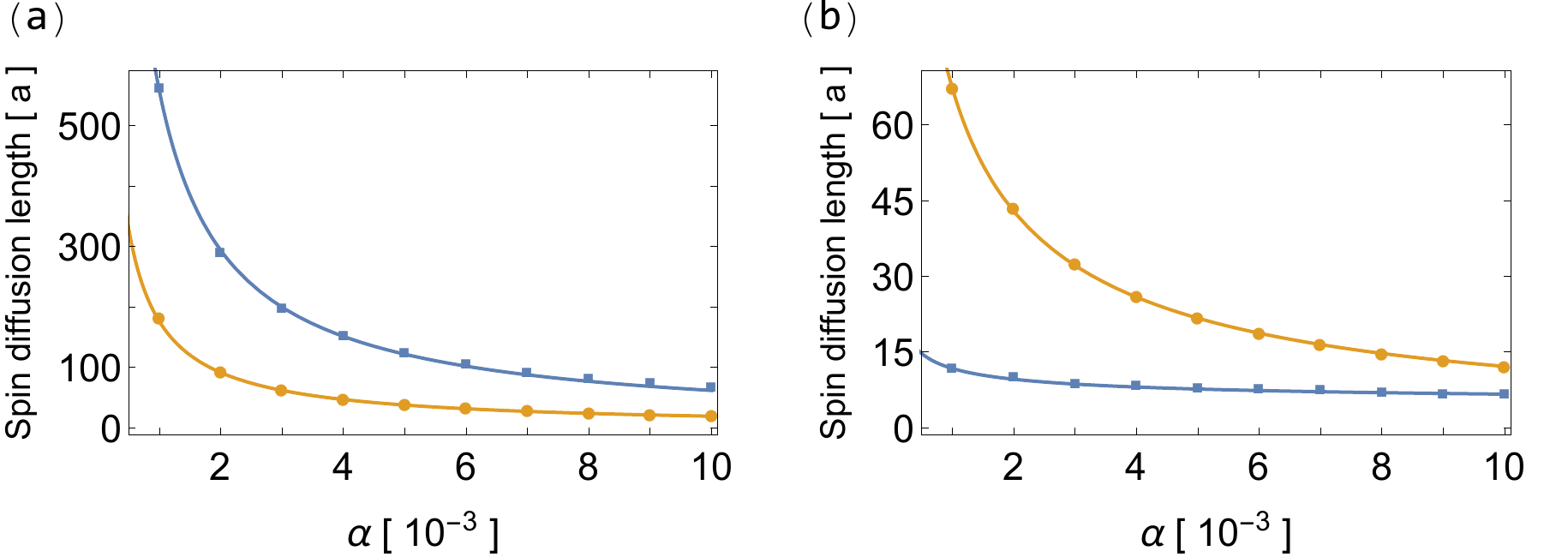}
\caption{\label{figS2}Spin diffusion lengths as a function of the bulk Gilbert damping $\alpha$,
for temperature $T=0.3\,J$ and average defect concentrations (a) $w=0$ and (b) $w=0.1$. 
Bulk and edge spin diffusion length are depicted in blue (squares) and yellow (circles) respectively.
The corresponding lines are fits with Eq.~(\ref{eq:length_alpha}).
For the bulk spin diffusion length in (b) it was necessary to add a small constant to the fitting function Eq.~(\ref{eq:length_alpha}).} 
\end{figure}
In this section we investigate the dependence of the bulk and edge spin diffusion lengths
introduced in Eq.~(\ref{eq:I_exp}) of the main text
on the (bulk) Gilbert damping $\alpha$.
Fig.~\ref{figS2} shows the bulk and edge diffusion lengths as function of the bulk Gilbert damping $\alpha$ for zero and finite defect concentrations $w$. 
Note that in the clean limit displayed in Fig.~\ref{figS2} (a),
the edge current decays faster than the bulk current for all damping values considered.
This is in agreement with the expectation that the high-frequency edge states are 
much more affected by the Gilbert damping than the low-frequency bulk magnons. 
On the other hand, as shown in Fig.~\ref{figS2} (b),
a finite defect concentration strongly suppresses the bulk diffusion length  
but only weakly influences the edge diffusion length.
Consequently, in the presence of defects, the edge current decays slower than the bulk current for virtually all values of Gilbert damping considered.  
In both cases, the damping dependence is fitted very well by 
\begin{equation} \label{eq:length_alpha}
\lambda^X(\alpha) = \frac{a}{\sqrt{ \gamma_1^X \alpha +\gamma_2^X \alpha^2 }},
\,\,\,
X = \textrm{bulk}, \textrm{edge} ,
\end{equation}
with $\gamma_1^X$ and $\gamma_2^X$ constants.
This may be understood phenomenologically by noting that
the spin diffusion length in ferromagnets is
$\lambda = v \sqrt{ \frac{2}{3} \tau_{\textrm{r}} \tau } $ \cite{Cornelissen16S},
where $v$ is the average magnon velocity, 
$\tau_{\textrm{r}}^{-1}$
is the magnon relaxation rate due to inelastic scattering,
and the total magnon relaxation rate is
$\tau^{-1} = \tau_{\textrm{r}}^{-1} + \tau_{\textrm{el}}^{-1}$
and also includes elastic scattering with defects and the sample boundaries
taken into account by $\tau_{\textrm{el}}^{-1}$.
As $\tau^{-1}_{\textrm{r}} \propto \alpha$ in ferromagnetic systems \cite{Cornelissen16S},
we obtain a spin diffusion length of the form of Eq.~(\ref{eq:length_alpha})
from this argument.

\end{widetext}


\begin{thebibliography}{99}
\bibitem{NoteSeparation}
In principle the edge states are connected continuously to the bulk bands, so our separation is only approximate.
However, since we consider a finite, mesoscopic system the energy levels are discrete, 
with only a small broadening due to the Gilbert damping.
We explicitly checked that there are no edge states with frequencies outside of the band gap,
so that the separation of the spin current into bulk and edge currents is essentially exact.
\end{thebibliography}
\end{document}